\documentclass[twocolumn,showpacs,superscriptaddress]{revtex4}
\usepackage{amssymb}
\usepackage{amsmath}
\usepackage{graphicx}
\usepackage{epsfig}
\usepackage{txfonts}
\usepackage{amsfonts}
\usepackage{CJK}

\begin{document}

\title{ Quantum anti-Zeno effect in artificial quantum systems }
\author{Qing Ai }
\affiliation{Institute of Theoretical Physics, Chinese Academy of
Sciences, Beijing, 100190, China}
\author{Jie-Qiao Liao }
\affiliation{Institute of Theoretical Physics, Chinese Academy of
Sciences, Beijing, 100190, China}
\author{C. P. Sun }
\affiliation{Institute of Theoretical Physics, Chinese Academy of
Sciences, Beijing, 100190, China}

\begin{abstract}
In this paper, we study a quantum anti-Zeno effect (QAZE) purely induced by
repetitive measurements for an artificial atom interacting with a
structured bath. This bath can be artificially realized with coupled
resonators in one dimension and possesses photonic band structure
like Bloch electron in a periodic potential. In the presence of
repetitive measurements, the pure QAZE is discovered as the observable
decay is not negligible even for the atomic energy level spacing
outside of the energy band of the artificial bath. If there were no
measurements, the decay would not happen outside of the band. In
this sense, the enhanced decay is completely induced by
measurements through the relaxation channels provided by the bath. Besides, we also discuss the controversial golden rule
decay rates originated from the van Hove's singularities and the
effects of the counter-rotating terms.
\end{abstract}
 \maketitle

\section{Introduction}

\label{app:introduction}

Generally speaking, the couplings between a quantum system and a bath leads
to the decay of the excitation in the system. The bath consists of many
harmonic oscillators with energy spectrum over a broad band. This
irreversible process requires bath's modes in approximate resonance with the
system's excited level. In other words, the approximately resonant modes
provide a relaxation channel for the decay of the system. Thus, this results
in a nonzero decay rate in the long run according to the Fermi golden rule~\cite{Breuer02,Yu94,Sun95}.

It was found out that such spontaneous decay could be usually suppressed by
frequent measurements~\cite{Khalhin68,Misra77,Koshinoa05}. This suppression phenomenon is called quantum Zeno
effect (QZE). But for some cases, the
above mentioned decay phenomena may be remarkably accelerated and thus the
quantum anti-Zeno effect (QAZE) occurs~\cite{Kofman00,Fischer01,Bar-Gill09}. In this case, the
measured decay rate is larger compared with the golden rule decay rate,
which is the decay rate purely resulting from the coupling to the bath in
the absence of the repetitive measurements.

This enhanced decay phenomenon is generally due to the associated effect of both the
coupling to the bath and the measurements. It depends on the matching
between the measurement's influence and the interacting spectrum. Now, a
question in point is whether measurements alone can induce decay. To answer this question,
we study the case with far off-resonant couplings in this paper. In other words, the
characteristic level spacing of the system lies far away from the energy
band of the bath. If there were no measurements, the decay could be
negligible. We will show that the QAZE indeed takes place on condition that
the bath provides a channel for energy relaxation.

By virtue of a specific example we illustrate the above
mentioned discovery. The total system is made up of an artificial atom and a
coupled-resonator waveguide with a narrow energy band, which was introduced
to investigate the coherent transport for single photon~\cite{Zhou08,Zhou09}%
. In this system, the level spacing of the artificial atom is feasibly
adjusted within and beyond the energy band of the bath. And the
coupled-resonator waveguide can be thought of as a structured bath with a
nonlinear dispersion relation. When the atomic transition frequency is tuned
beyond the energy band of the bath, the atomic decay is induced purely by
the measurement in contrast to the originally suppressed one. What is more
important, besides the bare excited state, our calculation also shows that
the QAZE exists for the physical excited state. This situation is different
from the case for the hydrogen atom where the QAZE does not occur for the
physical excited state~\cite{Zheng08} but the bare excited state~\cite{Ai10}%
. We emphasize that the reported phenomenon is a pure QAZE which is
repressed if no measurement is applied to the artificial atom.

In our consideration, starting from a general Hamiltonian without the
rotating-wave approximation (RWA)~\cite{Scully97}, we obtain an effective Hamiltonian by the
generalized Fr\"{o}hlich-Nakajima transformation and thus the effective
decay rate modified by the measurements. In general cases, the effect of the
counter-rotating terms can be omitted for it only leads to a small
correction to the atomic level spacing. But in some special cases, i.e.,
near the edges of the bath's energy band, the considerations with and
without the RWA seem to result in the opposite predictions of the atomic
decay. However, from an exact solution to the Schr\"{o}dinger equation
without the Wigner-Weisskopf approximation, we find out that there is no
singular behavior in this case.

The paper is structured as follows. In the next section, we describe the
total system including an artificial atom and a structured bath formed by a
coupled-resonator waveguide. With a generalized Fr\"{o}hlich-Nakajima
transformation, we obtain the effective Hamiltonian and thus the decay rate.
Moreover, we discuss the pure QAZE for both the bare excited state and the
physical excited state in Sec.~\ref{app:QAZE}. Since there may be singular
behavior for the decay phenomenon at the edges of the reservoir's energy
band, we analyze this situation from the exact solution to the Schr\"{o}%
dinger equation. Before a brief conclusion is summarized in Sec.~\ref%
{app:conclusion}, we present two proposals to put this model into practice
in Sec.~\ref{app:implementation}. Finally, in order to investigate the
singular behavior of the golden rule decay rate near the band edge, we offer
an exact solution to the Schr\"{o}dinger equation without the
Wigner-Weisskopf approximation in Appendix~\ref{app:appendix1}.

\section{Model Description}

\label{app:model}

\begin{figure}[tbp]
\includegraphics[scale=0.75]{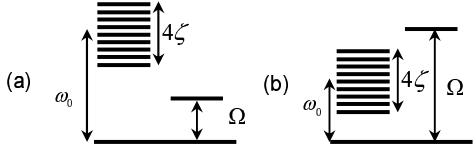}
\caption{The energy diagram of the total system. The level spacing of the
system is $\Omega$, while the energy spectrum of the bath is a band centered
at $\omega_0$ with width $4\zeta$. The level of the
system's excited state lies (a) below the lower limit of the band or (b) above the upper limit of the band.}
\label{setup}
\end{figure}

We consider a structured bath described by the Hamiltonian
\begin{equation}
H_{B}=\sum_{k}\omega _{k}b_{k}^{\dagger }b_{k}\text{,}
\end{equation}%
where $b_{k}$ and $b_{k}^{\dagger }$ are the annihilation and creation
operators of the $k$th mode, respectively. It possesses a nonlinear
dispersion relation
\begin{equation}
\omega _{k}=\omega _{0}-2\zeta \cos k\text{,}  \label{dispersionrelation}
\end{equation}%
which forms an energy band centered at $\omega _{0}$ with width $4\zeta $,
as shown in Fig.~\ref{setup}. Therein, we offer two situations that the atomic level spacing
is located beyond the energy band of the reservoir. To realize the above physical setup, we propose
two artificial architectures in circuit QED and photonic crystal plus quantum dot,
which will be shown explicitly in the Sec.~\ref{app:implementation}.

Besides, an artificial atom with level spacing $\Omega$ governed by the free Hamiltonian,
\begin{equation}
H_{S}=\frac{\Omega }{2}\sigma _{z}\text{,}
\end{equation}%
interacts with the structured bath. Here, $\sigma _{z}\equiv |e\rangle
\langle e|-|g\rangle \langle g|$ is the Pauli operator with $|e\rangle $ and$%
\ |g\rangle $ being the atomic excited and ground states, respectively. The
interaction Hamiltonian between the atom and the bath is given by
\begin{equation}
H_{I}=\sum_{k}g_{k}(b_{k}+b_{k}^{\dagger })(\sigma ^{+}+\sigma ^{-})\text{,}
\label{interactionH}
\end{equation}%
where $\sigma ^{+}=(\sigma ^{-})^{\dagger }\equiv |e\rangle \langle g|$ are
the raising and lowering operators for the atom, and we introduce the
coupling constants
\begin{equation}
g_{k}\equiv \frac{g}{\sqrt{N}},
\end{equation}%
which are equal for the $N$ modes. It should be emphasized that in the
interaction Hamiltonian~(\ref{interactionH}) we do not impose the RWA.

Thus the total system including the atom and the bath is governed by the
Hamiltonian
\begin{equation}
H=H_{S}+H_{B}+H_{I}.  \label{H}
\end{equation}

Due to the insolvability of the original Hamiltonian~(\ref{H}), we follow
the method introduced in Ref.~\cite{Ai10}, which is the generalized version~%
\cite{zhu-sun99} of the Fr\"{o}hlich-Nakajima transformation~\cite%
{Frohlich,Nakajima53}, to attain an effective Hamiltonian
\begin{eqnarray}
H_{\mathrm{eff}} &\simeq &e^{-S}He^{S}  \notag \\
&\simeq &H_{0}+H_{1}+\frac{1}{2}[H_{1},S]+\frac{1}{2}[H_{I},S],
\end{eqnarray}
where $H_{1}=H_{I}+[H_{0},S]$ is the first order term.

In order to eliminate the high-frequency terms $b_{k}^{\dagger }\sigma
^{+}+h.c.$, we choose the operator
\begin{equation}
S=\sum_{k}A_{k}(b_{k}^{\dagger }\sigma ^{+}-b_{k}\sigma ^{-})  \label{S}
\end{equation}%
with
\begin{equation}
A_{k}=-\frac{g_{k}}{\omega _{k}+\Omega }.
\end{equation}%
Then we obtain
\begin{equation}
H_{1}=\sum_{k}g_{k}(b_{k}\sigma ^{+}+h.c.)\text{.}
\end{equation}%
Further calculation shows that
\begin{eqnarray}
\lbrack H_{1},S] &=&-\sum_{k,k^{\prime }}A_{k^{\prime }}g_{k}(b_{k}^{\dagger
}b_{k^{\prime }}^{\dagger }+b_{k^{\prime }}b_{k})\sigma _{z}\text{,} \\
\lbrack H_{I},S] &=&-\sum_{k,k^{\prime }}A_{k^{\prime
}}g_{k}[(b_{k}^{\dagger }b_{k^{\prime }}+b_{k^{\prime }}^{\dagger
}b_{k})\sigma _{z}-2\sigma ^{-}\sigma ^{+}\delta _{kk^{\prime }}]  \notag \\
&&+[H_{1},S].
\end{eqnarray}

By omitting the high-frequency terms including $b_{k}^{\dagger }b_{k^{\prime
}}$ ($k\neq k^{\prime }$) and $b_{k}^{\dagger }b_{k^{\prime }}^{\dagger
}+b_{k^{\prime }}b_{k}$ in the above equations, we obtain%
\begin{eqnarray}
H_{\mathrm{eff}} &=&\sum_{k}\omega _{k}b_{k}^{\dagger }b_{k}+\frac{\Omega }{2%
}\sigma _{z}+\sum_{k}g_{k}(\sigma ^{+}b_{k}+h.c.)  \notag \\
&&-\sum_{k }A_{k}g_{k}b_{k}^{\dagger }b_{k }\sigma
_{z}+\sum_{k}A_{k}g_{k}\sigma ^{-}\sigma ^{+}.  \label{effectiveH}
\end{eqnarray}

For the case of single excitation, the fourth term on the right hand side of
Eq.~(\ref{effectiveH}) results in a small correction to the final
consequence and thus can be dropped off. In all, the effective Hamiltonian
is approximated as \newline
\begin{equation}
H_{\mathrm{eff}}=\sum_{k}\omega _{k}b_{k}^{\dagger }b_{k}+\frac{\Omega _{1}}{%
2}\sigma _{z}+\sum_{k}g_{k}(\sigma ^{+}b_{k}+h.c.)  \label{H_eff}
\end{equation}%
with the modified frequency
\begin{equation}
\Omega _{1}=\Omega -\sum_{k}A_{k}g_{k}\text{.}  \label{modifiedfrequency}
\end{equation}%
So far as the specific form of the interacting spectrum is concerned, the
modified atomic level spacing defined in Eq.~(\ref{modifiedfrequency}) is
\begin{eqnarray}
\Omega _{1} &=&\Omega +\sum_{k}\frac{g_{k}^{2}}{\omega _{k}+\Omega }  \notag
\\
&=&\Omega +\frac{N}{2\pi }\int_{-\pi }^{\pi }\frac{g_{k}^{2}}{\omega
_{k}+\Omega }dk  \notag \\
&=&\Omega +\frac{g^{2}}{\sqrt{(\omega _{0}+\Omega )^{2}-4\zeta ^{2}}}\text{.}
\end{eqnarray}%
By comparing Eq.~(\ref{H_eff}) with Eq.~(\ref{H}), we can see that the total
effect of the counter-rotating terms is to alter the atomic level spacing
while it leaves the coupling between the atom and the bath unchanged.

\section{Pure Anti-Zeno Effect}

\label{app:QAZE}

When we come to the QAZE, we refer to the survival probability of the atomic
excited state $|e\rangle $. In the previous studies, for the bare excited
state, we show that the QAZE still takes place in the presence of the
countering-rotating terms. However, the QAZE is erased and only is the QZE
left for the physical excited state. In this section, we discover that the
QAZE happens for both the two initial states in this artificial architecture.

\subsection{Quantum Anti-Zeno Effect for Spontaneous Decay}

In the following, we mainly focus on the QAZE for the spontaneous decay. In
other words, the total system is initially prepared in the bare excited
state $\left\vert e,\{0\}\right\rangle $, where the state $\left\vert
\{0\}\right\rangle \equiv |0_{1},\cdots ,0_{k},\cdots 0_{N}\rangle $ denotes
all of the bath's modes being in vacuum. Due to the specific transformation
of the form as Eq.~(\ref{S}), the initial state after the transformation $%
\exp (-S)\left\vert e,\{0\}\right\rangle =\left\vert e,\{0\}\right\rangle $
is unaltered. The survival probability of the atomic excited state coincides
with the one of the total system in its initial state \cite{Ai10}, i.e.,
\begin{eqnarray}
P_{e} &=&\text{Tr}_{B}\left( \left\vert e\right\rangle \left\langle
e\right\vert e^{-iHt}\left\vert e,\{0\}\right\rangle \left\langle
e,\{0\}\right\vert e^{iHt}\right)  \notag \\
&=&\left\vert \left\langle e,\{0\}\right\vert e^{-iH_{\text{eff}%
}t}\left\vert e,\{0\}\right\rangle \right\vert ^{2}\text{.}
\end{eqnarray}

As shown in Ref.~\cite{Ai10}, for the present case the decay rate after
repetitive measurements reads
\begin{equation}
R=2\pi \int\nolimits_{-\infty }^{\infty }d\omega F(\omega ,\Omega _{1},\tau
)G(\omega ),  \label{R}
\end{equation}%
which is an overlap integration of the level broadening induced by
measurements
\begin{equation}
F(\omega ,\Omega _{1},\tau )=\frac{\tau }{2\pi }\text{sinc}^{2}\left[ \frac{%
(\omega -\Omega _{1})\tau }{2}\right]
\end{equation}%
and the interacting spectrum
\begin{eqnarray}
G(\omega ) &=&\sum\nolimits_{k}g_{k}^{2}\delta (\omega -\omega _{k})  \notag
\\
&=&g_{k}^{2}\rho (\omega _{k})|_{\omega _{k}=\omega }.  \label{interspectrum}
\end{eqnarray}%
Also can the interacting spectrum be considered as the energy spectrum of
the bath weighed by the atomic coupling constants. In the following
discussions, we assume the resonator number $N$ to be such a large number
that it is reasonable to consider the density of state to \ be continuous in
the frequency space. Here, the density of state in the coupled-resonator
waveguide is
\begin{eqnarray}
\rho (\omega _{k}) &=&\frac{dn}{d\omega }  \notag \\
&=&\frac{N}{2\pi }\left\vert \frac{dk}{d\omega _{k}}\right\vert  \notag \\
&=&\frac{N}{2\pi }\left\vert \frac{1}{2\zeta \sin k}\right\vert  \notag \\
&=&\frac{N}{\pi }\frac{1}{\sqrt{4\zeta ^{2}-(\omega _{k}-\omega _{0})^{2}}},
\label{densityofstate}
\end{eqnarray}%
where $dn$ means the number of states within the frequency range $d\omega $.
Note that in the second line of the above equation, we have used the fact
that the distribution of the states in wavevector space is symmetric, with
the density $N/(2\pi )$. According to Eqs.~(\ref{dispersionrelation}) and~(%
\ref{densityofstate}), we know that the density of state $\rho (\omega _{k})$
has two singular points at both the ends of the band, as shown in Fig.~\ref%
{Gw}. And we can numerically verify that it fulfills the requirement for
normalization, i.e.,
\begin{equation}
\lim_{\xi \rightarrow 0}\int_{\omega _{0}-2\zeta +\xi }^{\omega _{0}+2\zeta
-\xi }\rho (\omega )d\omega =N\text{.}
\end{equation}%
Substitution of Eq.~(\ref{densityofstate}) for $\rho(\omega_k)$ in Eq.~(\ref%
{interspectrum}) leads to
\begin{equation}
G(\omega )=\frac{g^{2}}{\pi }\frac{1}{\sqrt{4\zeta ^{2}-(\omega -\omega
_{0})^{2}}}.
\end{equation}%
Clearly, this interacting spectrum is nonzero within the energy band of the
bath, ranging from $\omega _{0}-2\zeta$ to $\omega _{0}+2\zeta$, while
beyond the energy band it vanishes. This property of the interacting
spectrum can help us to understand the appearance of the pure QAZE.

\begin{figure}[ptb]
\includegraphics[bb=0 0 245 160,width=8 cm]{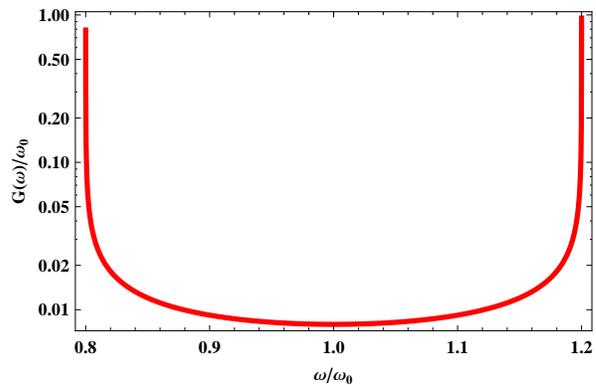}
\caption{ (color online) The interacting spectrum $G(\omega)$ vs $%
\omega$ with $\omega_0=1$, $g=0.1$, and $\zeta=0.1$.
The interacting spectrum is centered at $\omega_0$ $(=1)$ and with
width $4\zeta$ $(=0.4)$, i.e., extending from $0.8$ to $1.2$. There
are two singular points ($\omega/\omega_{0}=0.8$ and $1.2$)
at the two ends of the energy band.}
\label{Gw}
\end{figure}

When we refer to the QAZE, we make a comparison between the instantaneous
decay rate modified by the measurements and the unperturbed one, which is
the so-called golden rule decay rate $R_{G}$. By and large, the latter can
be obtained directly from the long time limit of Eq.~(\ref{R}), i.e., $R_{%
\mathrm{G}}=\lim_{\tau \rightarrow \infty }R$. In this case, since%
\begin{equation}
\lim_{\tau \rightarrow \infty }F(\omega ,\Omega _{1},\tau )=\delta (\omega
-\Omega _{1})\text{,}
\end{equation}%
the golden rule decay rate
\begin{equation}
R_{\mathrm{G}}=2\pi G(\Omega _{1})  \label{Rg}
\end{equation}%
is determined by the interacting spectrum at the modified atomic level
spacing.

\begin{figure}[tbp]
\includegraphics[bb=0 0 245 160,width=8 cm]{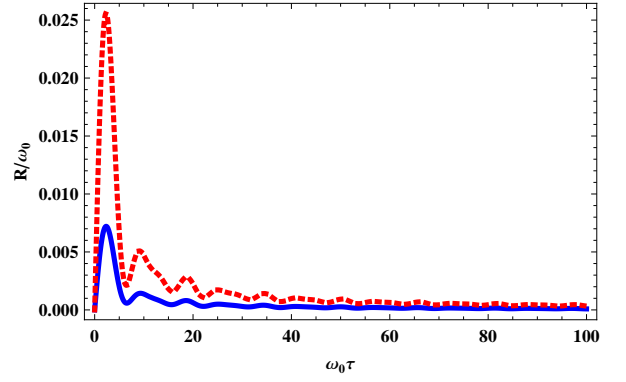}
\caption{(color online) The decay rate $R$ versus the scaled time $%
\omega _{0}\tau $ with $\omega _{0}=1$, $\zeta =0.1$%
, $g=0.1$, and $\Omega =2$. The blue solid line for the bare excited state
while the red dotted line for the physical excited state.}
\label{QAZE_R}
\end{figure}

Then, an interesting phenomena emerges. When there is no measurement applied
to the atom and the modified atomic level spacing is located outside of the
energy band of the bath, it is obvious that the excited atom will not decay
although it is coupled to the bath according to Eq.~(\ref{Rg}). It is a
comprehensible result since there is no energy level of the bath in
resonance with the atomic transition frequency. In other words, there is no
channel for the atomic excitation to relax. In this sense, the decay
phenomenon is greatly suppressed and therefore we obtain a zero golden rule
decay rate, i.e., $R_{\mathrm{G}}=2\pi G(\Omega _{1})=0$. The same physical
consequence could also be obtained from the Wigner-Weisskopf approximation
\cite{Sun03}. On the other hand, when the atom is measured, we calculate the
the decay rate $R$ according to Eq.~(\ref{R}) and find out that the decay
phenomenon could be observed due to the repetitive measurements no matter
how frequently the measurements are applied to atom. We remark that this is
a pure QAZE as the measurement-induced decay rate is definitely larger than
the vanishing golden rule decay rate. In Fig.~\ref{QAZE_R}, the
measurement-induced decay rate is plotted for different measurement
intervals. Therein, the level spacing of the atom is chosen as $\Omega =2$,
which is outside of the bath's energy band ranging from $0.8$ to $1.2$. It
is discovered that when we measure the atom repeatedly, the decay rate $R$
is nonzero. It is totally different from the golden rule decay rate $R_{%
\mathrm{G}}=0$ when there is no measurement applied to the atom. Therefore,
the QAZE is purely induced by the measurement.

This pure QAZE can be physically explained as follows. When the atom evolves
freely, there is only coupling between the atom and the bath. The excitation
originally in the atom can not relax to the bath since its energy level is
beyond the bath's energy band and thus there are no modes of the bath in
resonance with the atomic transition frequency. However, as the measurement
is applied, the inborn energy level is \textit{widely broadened}~\cite%
{Kofman00}. As long as there is overlap between the atomic broadened level
and the energy band of the bath, there exist channels for the atom to relax.
Therefore, the decay phenomenon comes into being. Mathematically, the
overlap integration~(\ref{R}) does not vanish in this case and thus results
in a nonzero decay rate.

\subsection{Anti-Zeno Effect for Physical Excited State}

In the previous subsection, the QAZE is displayed for the total system
initially prepared inthe bare excited state $\left\vert e,\{0\}\right\rangle
$. In Ref.~\cite{Loudon06}, it was announced that the bare ground state
should be replaced by the physical ground state $e^{S^{\prime }}\left\vert
g,\{0\}\right\rangle $ with the operator
\begin{equation}
S^{\prime }=\sum_{k}A_{k}(b_{k}^{\dagger }\sigma ^{-}-b_{k}\sigma
^{+}+b_{k}^{\dagger }\sigma ^{+}-b_{k}\sigma ^{-})\text{,}
\end{equation}%
due to the presence of the counter-rotating terms. Therefore, so far as the
initial state is concerned, the physical excited state $e^{S^{\prime
}}\left\vert e,\{0\}\right\rangle $\ substitutes for the bare exited state $%
\left\vert e,\{0\}\right\rangle $ \cite{Zheng08}. As a consequence, the QAZE
disappears and only is the QZE present for the physical excited state~\cite%
{Zheng08}. In this subsection, we will show that the QAZE still exists for
the physical excited state.

In this case, with respect to the physical excited state, the survival
probability of the atomic excited state after a projective measurement is%
\begin{eqnarray}
P_{e} &=&\text{Tr}_{B}\left( \left\vert e\right\rangle \left\langle
e\right\vert e^{-iHt}e^{S^{\prime }}\left\vert e,\{0\}\right\rangle
\left\langle e,\{0\}\right\vert e^{-S^{\prime }}e^{iHt}\right)  \notag \\
&=&\left\vert \left\langle e,\{0\}\right\vert e^{-iH_{\text{eff}}^{\prime
}t}\left\vert e,\{0\}\right\rangle \right\vert ^{2}\text{.}
\end{eqnarray}%
As shown in the above equation, the survival probability with respect to the
physical excited state under the original Hamiltonian~(\ref{H}) is
equivalent to the one with respect to the bare excited state under an
effective Hamiltonian
\begin{equation}
H_{\mathrm{eff}}^{\prime }=\sum_{k}\omega _{k}b_{k}^{\dagger }b_{k}+\frac{%
\Omega ^{\prime }}{2}\sigma _{z}+\sum_{k}g_{k}^{\prime }(\sigma
^{+}b_{k}+h.c.)
\end{equation}%
with a modified level spacing
\begin{equation}
\Omega ^{\prime }=\Omega +2\sum_{k}\frac{\Omega g_{k}A_{k}}{\omega
_{k}+\Omega }
\end{equation}%
and modified coupling constants%
\begin{equation}
g_{k}^{\prime }=\frac{2\Omega }{\omega _{k}+\Omega }g_{k}\text{.}
\end{equation}

Straightforward, we obtain the corresponding decay rate after $n$ repetitive
measurements%
\begin{equation}
R=2\pi \int\nolimits_{-\infty }^{\infty }d\omega F(\omega ,\Omega ^{\prime
},\tau )G^{\prime }(\omega )\text{,}
\end{equation}%
which is an overlap integration of the level broadening induced by
measurements centered at $\Omega ^{\prime }$
\begin{equation}
F(\omega ,\Omega ^{\prime },\tau )=\frac{\tau }{2\pi }\text{sinc}^{2}\left[
\frac{(\omega -\Omega ^{\prime })\tau }{2}\right]
\end{equation}%
and the modified interacting spectrum
\begin{eqnarray}
G^{\prime }(\omega ) &=&\sum\nolimits_{k}(g_{k}^{\prime })^{2}\delta (\omega
-\omega _{k})  \notag \\
&=&\frac{4\Omega ^{2}}{(\omega _{k}+\Omega )^{2}}g_{k}^{2}\rho (\omega
_{k})|_{\omega _{k}=\omega }  \notag \\
&=&\frac{4g^{2}\Omega ^{2}}{\pi (\omega +\Omega )^{2}\sqrt{4\zeta
^{2}-(\omega -\omega _{0})^{2}}}\text{.}
\end{eqnarray}

Also in the long limit, we obtain the corresponding golden rule decay rate%
\begin{equation}
R_{\mathrm{G}}=2\pi G(\Omega ^{\prime })\text{.}
\end{equation}%
On condition that the modified level spacing is beyond the band, the golden
rule decay rate vanishes similarly to the case with the bare excited state.
Thus, if there were nonvanishing decay phenomenon due to the measurements,
the pure QAZE would be observed. Yet, we plot the decay rate for this case
in Fig.~\ref{QAZE_R}. Notice that the measurement-induced decay rate for the
physical excited state is generally larger than the one for the bare exited
state.

\section{Decay Phenomenon near the Band Edge}

\label{app:DPBE}

As stated in the previous section, there are singular
points at both the ends of the bath's energy band. On account of the
discontinuous density of state at the edges of the band, we may justifiably
anticipate some exceptional phenomena around the band edge, especially the
ones due to the modification of the atomic level spacing. Generally
speaking, the difference between the modified atomic level spacing and the
original one is tiny small and thus can be neglected. However, for some
special cases, it seems to lead to totally opposite predictions about the
decay phenomenon induced by the coupling to the bath. We consider a specific
case when the level spacing of the artificial atom is tuned to the
neighborhood of the band edge, i.e., $\Omega <\omega _{0}+2\zeta $. If the
distance between the original atomic level spacing and the band edge is so
small that the modified level spacing is beyond the band. The theories with
the RWA and without the RWA offer opposite predictions about the decay
phenomenon, i.e., $R_{\mathrm{G}}^{\mathrm{RWA}}=2\pi G(\Omega )$ is nonzero
while $R_{\mathrm{G}}=2\pi G(\Omega ^{\prime })=0$ vanishes.

Besides, for the decay phenomenon exactly at the band edge, it seems that
there would be no atomic excited state existing as the golden rule decay
rate diverges due to the infinite large spectral density. Here, the
occurrence of singularities in the density of state is closely related to
the number of dimensions of the physical system~\cite{Hove53}.
Notwithstanding, all these controversies could be settled down if we resort
to the exact solution to the Schr\"{o}dinger equation, as shown in Appendix %
\ref{app:appendix1}.

The instantaneous decay rate without measurements is defined as
\begin{equation}
R(t)=-\frac{1}{\left\vert \alpha (t)\right\vert ^{2}}\frac{d\left\vert
\alpha (t)\right\vert ^{2}}{dt}\text{,}
\end{equation}
where the survival probability of the atomic excitation is
\begin{equation}
\left\vert \alpha (t)\right\vert ^{2}=\left\vert
A_{1}e^{p_{1}t}+A_{2}e^{p_{2}t}+\int\limits_{-2\zeta }^{2\zeta }C(x)e^{i(%
\frac{\Omega ^{\prime }}{2}-\omega _{0}+x)t}dx\right\vert ^{2}
\end{equation}
and its rate of change is
\begin{eqnarray}
\frac{d\left\vert \alpha (t)\right\vert ^{2}}{dt} &=&2\Re (I_{1}\times
I_{2}^{\ast })  \notag \\
&&-2A_{1}A_{2}(ip_{1}-ip_{2})\sin (ip_{1}-ip_{2})t  \notag \\
&&+2\Re \lbrack (A_{1}p_{1}e^{p_{1}t}+A_{2}p_{2}e^{p_{2}t})I_{1}^{\ast }]
\notag \\
&&+2\Re \lbrack (A_{1}e^{p_{1}t}+A_{2}e^{p_{2}t})I_{2}^{\ast }]\text{,}
\end{eqnarray}
where the integrals are defined as
\begin{eqnarray}
I_{1} &=&\int\limits_{-2\zeta }^{2\zeta }C(x)e^{i(\frac{\Omega ^{\prime }}{2}
-\omega _{0}+x)t}dx\text{,} \\
I_{2} &=&i\int\limits_{-2\zeta }^{2\zeta }C(x)\left(\frac{\Omega ^{\prime }}{%
2}-\omega _{0}+x\right)e^{i(\frac{\Omega ^{\prime }}{2} -\omega _{0}+x)t}dx%
\text{,}
\end{eqnarray}
with
\begin{equation}
C(x)=\frac{1}{\pi }\frac{g^{2}\sqrt{4\zeta ^{2}-x^{2}}}{(4\zeta
^{2}-x^{2})(\Omega ^{\prime }-\omega _{0}+x)^{2}+g^{4}}\text{.}
\end{equation}
The sign $\Re (x)$ means the real part of $x$ and the coefficients are given
as
\begin{equation}
A_{j}=\frac{[(ip_{j}+\frac{\Omega ^{\prime }}{2}-\omega _{0})^{2}-4\zeta
^{2}]}{[(ip_{j}+\frac{\Omega ^{\prime }}{2}-\omega _{0})^{2}-4\zeta
^{2}]+(ip_{j}-\frac{\Omega ^{\prime }}{2})(ip_{j}+\frac{\Omega ^{\prime }}{2}
-\omega _{0})}\text{.}
\end{equation}
And $p_{1}$ and $p_{2}$ are the two solutions to
\begin{equation}
\left(ip-\frac{\Omega ^{\prime }}{2}\right)^{2}\left[\left(ip+\frac{\Omega
^{\prime }}{2}-\omega _{0}\right)^{2}-4\zeta ^{2}\right]+g^{4}=0,
\end{equation}
with
\begin{eqnarray}
&&ip_{1}+\frac{\Omega ^{\prime }}{2}-\omega _{0}>2\zeta, \\
&&ip_{2}+\frac{\Omega ^{\prime }}{2}-\omega _{0}<-2\zeta.
\end{eqnarray}

\begin{figure}[tbp]
\includegraphics[bb=0 0 250 160,width=8 cm]{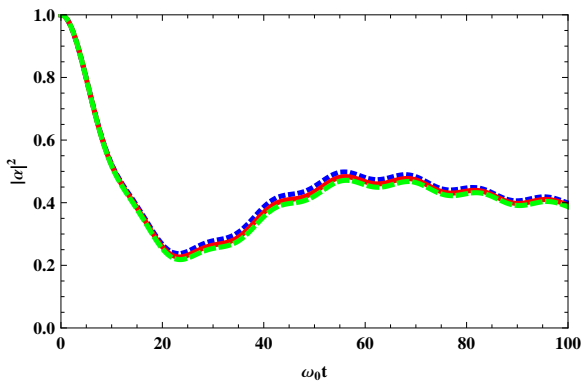}
\caption{ (color online) The survival probability $|\alpha |^{2}$
near the band edge with $\omega _{0}=1$, $g=0.1$ and $\zeta %
=0.1$. The blue dotted line for the modified frequency $\Omega ^{\prime
}=1.203$ and the green dashed line for the original frequency $\Omega =1.198$%
. And the red solid line is just the case of the atomic level spacing
exactly at the band edge, i.e., $\Omega =1.2$. }
\label{alpha3}
\end{figure}

In order to show the above result explicitly, we plot the free evolution of
the atomic excitation around the band edge in Fig.~\ref{alpha3}. It is seen
that the initial atomic excitation will nonexceptionally decay into a steady
value for the three cases, of which the atomic level spacings are
distributed within the band, beyond the band and exactly at the band edge.
Here, the original and modified atomic level spacing are tuned to the either
side of the band edge. And the differences among the survival probabilities
are negligible as the three frequencies are nearly identical.

\begin{figure}[ptb]
\includegraphics[bb=0 0 250 160,width=8 cm]{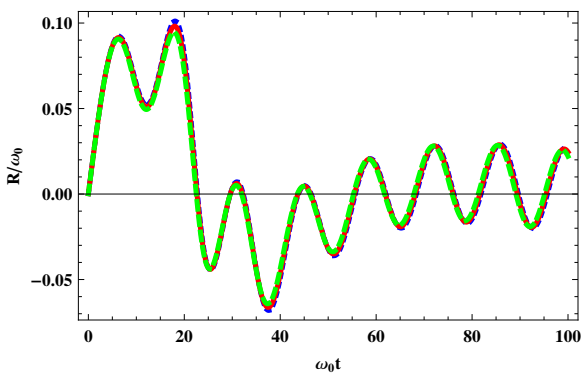}
\caption{ (color online) The instantaneous decay rate $R$ near the band edge
with the same parameters given in Fig. \ref{alpha3}. Notice that
three curves are nearly overlap. }
\label{R3}
\end{figure}

Besides, Fig.~\ref{R3} presents the instantaneous decay rate for the above
situations. It is seen that despite some oscillation around zero, the decay
rate $R$ always remains finite no matter whether the level spacing is at the
band edge or not. And the differences between them is so small that we can
neglect them. Further investigation shows the survival probability tends to
be a steady value after an initial decay. Here, we emphasize that the
divergent golden rule decay rate at the band edge is due to the improper
Wigner-Weisskopf approximation made in the deduction. In this case, the
spectral density varies sharply around the edge of the band. Since the
atomic excitation decays into all of the channels around the atomic
frequency, we shall average all the contributions from these channels
instead of counting on the single one which exactly equals to the atomic
frequency. Intuitionally, the decay rate for the atomic frequency at the
band edge does not diverge.

\section{Physical implementations of the artificial system}

\label{app:implementation}

In this section, we propose two possible physical setups to observe the
above phenomena. As mentioned above, the artificial system is composed of a
tunable two-level system and a coupled-resonator waveguide. Therefore, the
primal requirement for physical implementation is to provide a
coupled-resonator array and a tunable two-level system, which is coherently
coupled to one of the resonators in the waveguide. Currently, there are
several potential candidates. For instance, in superconducting circuit QED,
coupled superconducting transmission line resonator array can be realized to
interact with a superconducting charge qubit. And in semiconductor microwave
cavity QED, coupled photonic crystal cavity array interacts with an
artificial atom formed by a semiconductor quantum dot. In the following
subsections we will address the two systems respectively.

\subsection{Circuit QED}

\begin{figure}[tbp]
\includegraphics[bb=70 490 566 780,width=7
cm]{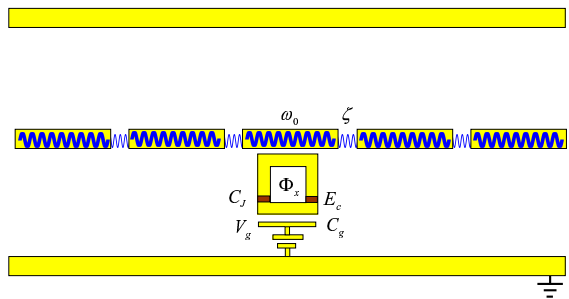}
\caption{ (color online) Implementation in circuit QED. The total system
consists of an artificial atom and a structured bath formed by a coupled
transmission line resonator waveguide. The atom is a superconducting charge
qubit, which is located at the zeroth resonator. }
\label{scheme1}
\end{figure}

First of all, we consider the artificial system to be realized in the
circuit QED~\cite{Blais04} as shown in Fig.~\ref{scheme1}.
The artificial atom is a Cooper pair box, also called charge qubit, which is
a dc current superconducting quantum interference device. It is a
superconducting island connected to two Josephson junctions. Around the
degenerate point, the Cooper pair box is approximated as a two-level system
with level spacing
\begin{equation}
\Omega =\sqrt{B_{x}^{2}+B_{z}^{2}}\text{.}
\end{equation}%
And the two eigen states are defined as
\begin{eqnarray}
\left\vert g\right\rangle &=&\sin \left( \theta /2\right) \left\vert
0\right\rangle +\cos \left( \theta /2\right) \left\vert 1\right\rangle \text{%
,} \\
\left\vert e\right\rangle &=&\cos \left( \theta /2\right) \left\vert
0\right\rangle -\sin \left( \theta /2\right) \left\vert 1\right\rangle \text{%
,}
\end{eqnarray}%
where $\left\vert 0\right\rangle $ and $\left\vert 1\right\rangle $ denote
the states with $0$ and $1$ extra Cooper pair on the island, respectively.
Here, we also introduce the mixing angle
\begin{equation}
\theta =\tan ^{-1}\left( \frac{B_{x}}{B_{z}}\right) .
\end{equation}%
On one hand, the level spacing $\Omega $ is tunable since the energy
\begin{equation}
B_{x}=4E_{c}(2n_{g}-1)\text{,}
\end{equation}%
which originates from the charging energy of
\begin{equation}
n_{g}=\frac{C_{g}V_{g}}{2e}
\end{equation}%
extra Cooper pair on the island, can be varied by changing the gate voltage $%
V_{g}$ applied to the gate capacitor $C_{g}$. Here, $C_{J}$ is the
capacitance of the single Josephson junction, and
\begin{equation}
E_{c}=\frac{e^{2}}{2(C_{g}+2C_{J})}\text{.}
\end{equation}%
On the other hand, the level spacing can also be adjusted as the energy
\begin{equation}
B_{z}=2E_{J}\cos \left( \frac{\pi \Phi _{x}}{\Phi _{0}}\right)
\end{equation}%
is induced by the controllable applied magnetic flux $\Phi _{x}$, where $%
E_{J}$ is the Josephson energy and $\Phi _{0}$ is the flux quanta.

In addition, a coplanar transmission line resonator is cut into $N$ pieces
to form a coupled-resonator waveguide~\cite{liao}. And the coupling constant
$\zeta $ between two neighboring resonators is determined by the coupling
mechanism. Placed at the antinode of single-mode electromagnetic field, the
atom only interacts with the electric field with the coupling strength to be
\begin{equation}
g=\frac{eC_{g}\sin \theta }{C_{g}+2C_{J}}\sqrt{\frac{\omega _{0}}{Lc}},
\end{equation}%
where $\omega _{0}$ is the frequency of the single mode in the transmission
line with length $L$ and capacitance per unit length $c$. For the
experimentally accessible parameters, we have $\omega _{0}\sim 5-10$ GHz and
$\Omega \sim 5-15$ GHz \cite{Wallraff04}. Therefore, the above mentioned
parameters are realizable in practice.

\subsection{Photonic Crystal plus Quantum Dot}

\begin{figure}[tbp]
\includegraphics[bb=80 560 525 750,width=7
cm]{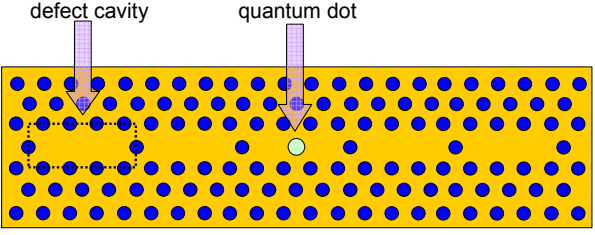}
\caption{ (color online) Implementation in the photonic crystal. The system
is made up of a photonic crystal cavity fabricated in a GaAs membrane
containing a central layer with self-assembled InGaAs quantum dot inside.
The quantum dot plays the part as an artificial atom. }
\label{scheme2}
\end{figure}

In addition to the circuit QED, the above mentioned system can also be
realized in the photonic crystal. As shown in Fig.~\ref{scheme2}, a two
dimensional photonic crystal is fabricated in a sandwich-like architecture.
The crystal consists of a square lattice of high-index dielectric rods. We
attain a defected cavity by removing three rods. And the coupled defected
cavities form the artificial bath, while the quantum dot within the central
layer plays the role as the artificial atom. The strong coupling between a
quantum dot and a single cavity was experimentally realized~\cite%
{Akahane03,Yoshie04,Englund10}. Besides, multiple coupled photonic crystal
cavities have been already achieved to show all-optical electromagnetically
induced transparency~\cite{Yang09}. And the model of coupled cavities in the
photonic crystal was put forward to investigate theoretically photonic
Feshbach resonance~\cite{Xu08}.

\section{Conclusion and Discussions}

\label{app:conclusion}

To summarize, we investigate the enhanced decay phenomenon in the total system
composed of an artificial atom interacting with a structured bath. We apply
a generalized Fr\"{o}hlich-Nakajima transformation to obtain the effective
Hamiltonian without the use of the RWA. It is discovered that the originally suppressed
decay is enhanced due to the frequent projective measurements when the
atomic frequency is tuned beyond the energy band of the reservoir. And the
QAZE is present not only for the bare excited state but also for the
physical excited state. This is different from the case for the hydrogen
atom where the QAZE only exists for the former. We also remark that this is
a pure QAZE entirely resulting from the measurement-induced atomic level
broadening. Besides, we also discuss the singular behavior of the golden
rule decay rate near the band edge. Without the use of Wigner-Weisskopf
approximation, we attain the exact form of the unperturbed decay rate. It is
found out that despite the oscillations the decay rate without measurements
always remains finite in contrast to the divergent golden rule decay rate at
the band edge. In addition, when the atomic frequency is tuned outside of
the band, the exact decay rate tends to vanish in the long run, which is in
accordance with the one obtained with the Wigner-Weisskopf approximation.

However, there are still some problems remaining. Generally speaking, the
QAZE refers to the specific case where the measurement-induced decay rate is
faster than the unperturbed one, also called golden rule decay rate. In
other cases, someone made a comparison between the decay phenomenon
disturbed by the repetitive measurements and the free evolution, i.e.,~\cite%
{Lizuain09}. Besides, the interaction between the artificial bath and its
environment broadens its energy band and thus the coupling spectrum.
Therefore, the golden rule decay rate may not vanish due to the possible
nonzero coupling spectrum at the atomic level spacing although it can be
initially tuned outside of the bath's energy band. What is more important,
the measurement used here is considered as an ideal projection. In some cases,
by optically pumped into an auxiliary level, the population of the concerned level
in the dynamical evolution was measured~\cite{Itano90}. In the near future,
we may study the QAZE for this case.

We would like to express his gratitude towards T. Shi and Y. Li for many
stimulating discussions. This work was supported by NSFC through grants
10974209 and 10935010 and by the National 973 program (Grant
No.~2006CB921205).

\appendix

\section{Unperturbed Decay Rate Without Wigner-Weisskopf Approximation}

\label{app:appendix1}

In this Appendix, we present an exact solution for the excited state
population of a two-level atom coupled with a coupled-resonator waveguide.
This problem is equivalent to the spontaneous emission of an artificial atom
interacting with a structured bath. Without Wigner-Weisskopf approximation,
we obtain the exact solution to the Schr\"{o}dinger equation by the method
of Laplace transform.

The total system under our consideration is governed by the Hamiltonian
\begin{equation}
H=\sum_{k}\omega _{k}b_{k}^{\dagger }b_{k}+\frac{\Omega }{2}\sigma
_{z}+\sum_{k}g_{k}(b_{k}\sigma ^{+}+b_{k}^{\dagger }\sigma ^{-})\text{,}
\end{equation}%
where $\Omega ^{\prime }$ is replaced by $\Omega $ for ease of notation.
Here the dispersion relation is
\begin{equation}
\omega _{k}=\omega _{0}-2\zeta \cos k
\end{equation}%
with $k\in (-\pi ,\pi ]$ and the coupling constants for all modes are equal
as
\begin{equation}
g_{k}=\frac{g}{\sqrt{N}}.
\end{equation}%
Since the total excitation number operator $\sum_{k}b_{k}^{\dagger
}b_{k}+|e\rangle \langle e|$ of the system is a conservable quantity, we can
express a general wavefunction, in single-excitation space, of the system at
time $t$ as
\begin{equation}
\left\vert \Psi (t)\right\rangle =\alpha (t)\left\vert e,\{0\}\right\rangle
+\sum\limits_{k}\beta _{k}(t)\left\vert g,1_{k}\right\rangle ,
\end{equation}%
where $\left\vert 1_{k}\right\rangle \equiv |0_{1},\cdots ,1_{k},\cdots
0_{N}\rangle $ denotes that the $k$th mode possesses a single photon while
other modes are in vacuum. By comparing the coefficients on the both sides
of the Schr\"{o}dinger equation
\begin{equation}
i\partial _{t}\left\vert \Psi (t)\right\rangle =H\left\vert \Psi
(t)\right\rangle \text{,}
\end{equation}%
we have
\begin{eqnarray}
i\dot{\alpha}(t) &=&\frac{\Omega }{2}\alpha (t)+\sum\limits_{k}g_{k}\beta
_{k}(t)\text{,} \\
i\dot{\beta}_{k}(t) &=&\left( \omega _{k}-\frac{\Omega }{2}\right) \beta
_{k}(t)+g_{k}\alpha (t)\text{.}
\end{eqnarray}

By making Laplace transform
\begin{equation}
\tilde{\alpha}(p)=\int\limits_{0}^{\infty }dt\alpha (t)e^{-pt}
\end{equation}%
and by virtue of
\begin{equation}
\int\limits_{0}^{\infty }dt\dot{\alpha}(t)e^{-pt}=p\tilde{\alpha}(p)-\alpha
(0)\text{,}
\end{equation}%
we obtain
\begin{eqnarray}
\tilde{\alpha}(p) &=&\frac{i\alpha (0)}{\frac{\Omega }{2}-ip+\sum\limits_{k}%
\frac{g_{k}^{2}}{ip+(\frac{\Omega }{2}-\omega _{k})}}  \notag \\
&=&\frac{1}{p+i\frac{\Omega }{2}+\sum\limits_{k}\frac{g_{k}^{2}}{p\newline
+i(\omega _{k}-\frac{\Omega }{2})}}\text{,}  \label{AlphaP}
\end{eqnarray}%
where we have used the initial condition
\begin{equation}
\alpha (0)=1,\hspace{0.5cm}\beta _{k}(0)=0\text{.}
\end{equation}

\bigskip When the atomic level spacing is far off-resonant with all bath's
modes, i.e.,
\begin{equation}
\Omega >>\omega _{0}+2\zeta \text{,}
\end{equation}%
$p$ in the third term of the denominator on the right hand side of Eq. (\ref%
{AlphaP}) can be approximated as $-i\Omega /2$, namely the Wigner-Weisskopf
approximation. By means of the inverse Laplace transformation, we have a
nonvanishing atomic excitation amplitude%
\begin{equation}
\alpha (t)=e^{-i(\frac{\Omega }{2}+\frac{g^{2}}{\sqrt{(\omega _{0}-\Omega
)^{2}-4\zeta ^{2}}})t}\text{.}
\end{equation}%
All the effect of the coupling to the bath contributes an additional phase.

In the following, we will show the exact solution of $\alpha (t)$\ since the
above used Wigner-Weisskopf approximation may fail for the cases that there
are modes approximately in resonance with the atomic excited level. In order
to calculate $\alpha (t)$, we need to calculate the inverse Laplace
transform of $\tilde{\alpha}(p)$. Therefore we shall find out the branch cut
and poles of $\tilde{\alpha}(p)$ at first. The branch cut is defined as the
line of which the limits on the two sides are different from each other,
i.e.,
\begin{equation}
p\in \left[ i\left( \Omega /2-\omega _{0}-2\zeta \right) ,i\left( \Omega
/2-\omega _{0}+2\zeta \right) \right] \text{.}
\end{equation}%
The poles can be found out directly from
\begin{equation}
p+i\frac{\Omega }{2}+\sum\limits_{k}\frac{g_{k}^{2}}{p\newline
+i(\omega _{k}-\frac{\Omega }{2})}=0\text{.}  \label{SP}
\end{equation}%
The second term on the left hand side of the above equation can be expressed
as
\begin{eqnarray}
\sum\limits_{k}\frac{g_{k}^{2}}{p\newline
+i(\omega _{k}-\frac{\Omega }{2})} &=&\frac{N}{2\pi }\int\limits_{-\pi
}^{\pi }dk\frac{g_{k}^{2}}{p\newline
+i(\omega _{k}-\frac{\Omega }{2})}  \notag \\
&=&\frac{g^{2}}{2\pi }\int\limits_{-\pi }^{\pi }dk\frac{1}{p-i(\frac{\Omega
}{2}-\omega _{0}+2\zeta \cos k)}  \notag \\
&=&\frac{g^{2}}{2\pi \zeta }\oint\limits_{\left\vert z\right\vert =1}\frac{dz%
}{iz}\frac{1}{\frac{p-i(\frac{\Omega }{2}-\omega _{0})}{\zeta }-i(z+\frac{1}{%
z})}  \notag \\
&=&\frac{g^{2}}{2\pi \zeta }\oint\limits_{\left\vert z\right\vert =1}\frac{dz%
}{z^{2}+\frac{ip+\frac{\Omega }{2}-\omega _{0}}{\zeta }z+1}  \notag \\
&=&\frac{g^{2}}{2\pi \zeta }\oint\limits_{\left\vert z\right\vert =1}\frac{dz%
}{z^{2}+Mz+1}
\end{eqnarray}%
with
\begin{equation}
M=\frac{ip+\frac{\Omega }{2}-\omega _{0}}{\zeta }\text{.}
\end{equation}%
Obviously, there are two solutions
\begin{equation}
z_{\pm }=\frac{-M\pm \sqrt{M^{2}-4}}{2}
\end{equation}%
for equation
\begin{equation}
z^{2}+Mz+1=0.
\end{equation}

In the case of
\begin{equation}
M>2
\end{equation}
or equivalently
\begin{equation}
ip+\frac{\Omega }{2}-\omega _{0}>2\zeta \text{,}  \label{Condition1}
\end{equation}
we have
\begin{equation}
0>z_{+}>-1,
\end{equation}
which is within the integration loop and
\begin{equation}
z_{-}<-1,
\end{equation}
which is outside of the integral loop. Therefore,
\begin{eqnarray}
\sum\limits_{k}\frac{g_{k}^{2}}{p\newline
+i(\omega _{k}-\frac{\Omega }{2})} &=&\frac{g^{2}}{2\pi i\zeta }
\oint\limits_{\left\vert z\right\vert =1}\frac{dz}{z^{2}+Mz+1}  \notag \\
&=&\frac{g^{2}}{2\pi \zeta }2\pi i\lim_{z\rightarrow z_{+}}\frac{(z-z_{+})}{
(z-z_{+})(z-z_{-})}  \notag \\
&=&\frac{ig^{2}}{\zeta (z_{+}-z_{-})}  \notag \\
&=&\frac{ig^{2}}{\zeta \sqrt{M^{2}-4}}  \notag \\
&=&\frac{ig^{2}}{\sqrt{(ip+\frac{\Omega }{2}-\omega _{0})^{2}-4\zeta ^{2}}}
\text{.}
\end{eqnarray}
By substituting the above equation into Eq.~(\ref{SP}), we attain
\begin{equation}
ip-\frac{\Omega }{2}-\frac{g^{2}}{\sqrt{(ip+\frac{\Omega }{2}-\omega
_{0})^{2}-4\zeta ^{2}}}=if_{1}(p)=0\text{.}  \label{BS1}
\end{equation}
Notice that the solution to the above equation is $p_{1}$ which should
fulfill the requirement in Eq.~(\ref{Condition1}).

Similarly, when
\begin{equation}
M<-2
\end{equation}
or equivalently
\begin{equation}
ip+\frac{\Omega }{2}-\omega _{0}<-2\zeta,  \label{Condition2}
\end{equation}
we have
\begin{equation}
0<z_{-}<1
\end{equation}
within the integral loop while
\begin{equation}
z_{+}>1
\end{equation}
outside of the integral loop. Therefore,
\begin{equation}
\sum\limits_{k}\frac{g_{k}^{2}}{p\newline
+i(\omega _{k}-\frac{\Omega }{2})}=\frac{-ig^{2}}{\sqrt{(ip+\frac{\Omega }{2}
-\omega _{0})^{2}-4\zeta ^{2}}}\text{.}
\end{equation}
Therefore, we obtain the equation for the other singular point $p_{2}$,
i.e.,
\begin{equation}
ip-\frac{\Omega }{2}+\frac{g^{2}}{\sqrt{(ip+\frac{\Omega }{2}-\omega
_{0})^{2}-4\zeta ^{2}}}=if_{2}(p)=0\text{,}  \label{BS2}
\end{equation}
which should fulfill the requirement in Eq.~(\ref{Condition2}).

\begin{figure}[ptb]
\includegraphics[scale=0.5]{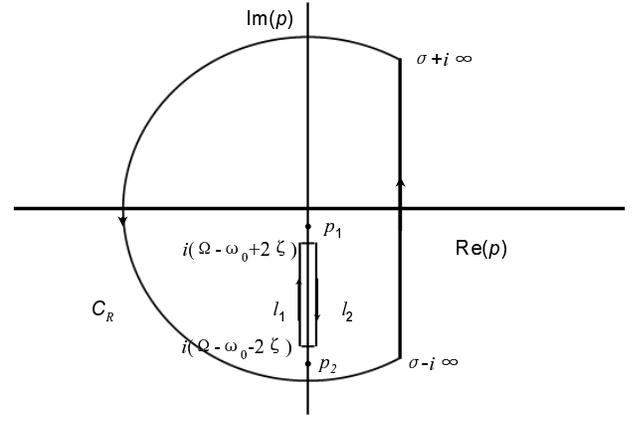}
\caption{ Integration path for Eq.~(\ref{Loop}).}
\label{p}
\end{figure}

In the following, we will calculate $\alpha (t)$ by making use of the
inverse Laplace transform,
\begin{equation}
\alpha (t)=\frac{1}{2\pi i}\int\limits_{\sigma -i\infty }^{\sigma +i\infty
}dp\tilde{\alpha}(p)e^{pt}.
\end{equation}
As shown in Fig.~\ref{p}, the contour integration is divided into four parts
as follows
\begin{equation}
\int\limits_{\sigma -i\infty }^{\sigma +i\infty
}+\int\limits_{C_{R}}+\int\limits_{l_{1}}+\int\limits_{l_{2}}=\oint
=\sum_{j} \mathrm{res}[\tilde{\alpha}(p_{j})e^{p_{j}t}]\text{,}  \label{Loop}
\end{equation}
where $\mathrm{res}[F(p)]$ denotes the residue of function $F(p)$ at $p$.
Thus, we have
\begin{eqnarray}
\alpha (t) &=&\frac{1}{2\pi i}\int\limits_{\sigma -i\infty }^{\sigma
+i\infty }dp\tilde{\alpha}(p)e^{pt}  \notag \\
&=&\sum_{j}\mathrm{res}[\tilde{\alpha}(p_{j})e^{p_{j}t}]-\int
\limits_{C_{R}}-\int\limits_{l_{1}}-\int\limits_{l_{2}}  \notag \\
&=&\sum_{j}\mathrm{res}[\tilde{\alpha}(p_{j})e^{p_{j}t}]-\int
\limits_{l_{1}}-\int\limits_{l_{2}}\text{,}
\end{eqnarray}
where we have used the generalized Jordan Lemma \cite{Liang98}
\begin{equation}
\int\limits_{C_{R}}=0.
\end{equation}

For the pole given by Eqs.~(\ref{Condition1}) and~(\ref{BS1}), the residue
is given as
\begin{equation}
\mathrm{res}[\tilde{\alpha}(p_{1})e^{p_{1}t}]=\frac{e^{p_{1}t}}{\frac{%
df_{1}(p)}{dp}|_{p=p_{1}}}=A_{1}e^{p_{1}t}\text{,}
\end{equation}
where
\begin{equation}
A_{1}=\frac{[(ip_{1}+\frac{\Omega }{2}-\omega _{0})^{2}-4\zeta ^{2}]}{
[(ip_{1}+\frac{\Omega }{2}-\omega _{0})^{2}-4\zeta ^{2}]+(ip_{1}-\frac{%
\Omega }{2})(ip_{1}+\frac{\Omega }{2}-\omega _{0})}\text{.}  \label{A1}
\end{equation}

For the singular point given by Eq.~(\ref{BS2}) and Eq.~(\ref{Condition2}),
the residue is given as
\begin{equation}
\mathrm{res}[\alpha (p_{2})e^{p_{2}t}]=\frac{e^{p_{2}t}}{\frac{df_{2}(p)}{dp}
|_{p=p_{2}}}=A_{2}e^{p_{2}t}\text{,}
\end{equation}
where
\begin{equation}
A_{2}=\frac{[(ip_{2}+\frac{\Omega }{2}-\omega _{0})^{2}-4\zeta ^{2}]}{
[(ip_{2}+\frac{\Omega }{2}-\omega _{0})^{2}-4\zeta ^{2}]+(ip_{2}-\frac{%
\Omega }{2})(ip_{2}+\frac{\Omega }{2}-\omega _{0})}.  \label{A2}
\end{equation}

In the following we will calculate the contribution from the branch cut as
\begin{eqnarray}
-\int\limits_{l_{1}}-\int\limits_{l_{2}} &=&-\frac{1}{2\pi i}
[\int\limits_{ip_{\min }}^{ip_{\max }}\frac{e^{pt}dp}{(p+i\frac{\Omega }{2}
)+\sum\limits_{k}\frac{g_{k}^{2}}{p\newline
+i(\omega _{k}-\frac{\Omega }{2})-0^{+}}}  \notag \\
&&+\int\limits_{ip_{\max }}^{ip_{\min }}\frac{e^{pt}dp}{(p+i\frac{\Omega }{2}
)+\sum\limits_{k}\frac{g_{k}^{2}}{p-i(\frac{\Omega }{2}-\omega _{k})+0^{+}}}]
\notag \\
&=&-\frac{1}{2\pi i}[\int\limits_{p_{\min }}^{p_{\max }}\frac{e^{ipt}dp}{(p+
\frac{\Omega }{2})-\sum\limits_{k}\frac{g_{k}^{2}}{p-(\frac{\Omega }{2}
-\omega _{k})+i0^{+}}}  \notag \\
&&+\int\limits_{p_{\max }}^{p_{\min }}\frac{e^{ipt}dp}{(p+\frac{\Omega }{2}
)-\sum\limits_{k}\frac{g_{k}^{2}}{p-(\frac{\Omega }{2}-\omega _{k})-i0^{+}}}%
] \text{,}  \label{BC}
\end{eqnarray}
where the limits for the integral are%
\begin{eqnarray}
p_{\min } &=&\frac{\Omega }{2}-\omega _{0}-2\zeta \text{,} \\
p_{\max } &=&\frac{\Omega }{2}-\omega _{0}+2\zeta \text{.}
\end{eqnarray}
In the denominator of Eq.~(\ref{BC}),
\begin{eqnarray}
&&\sum\limits_{k}\frac{g_{k}^{2}}{p-(\frac{\Omega }{2}-\omega _{k})\pm
i0^{+} }  \notag \\
&=&\frac{N}{2\pi }\int\limits_{-\pi }^{\pi }\frac{g_{k}^{2}dk}{p-(\frac{
\Omega }{2}-\omega _{k})\pm i0^{+}}  \notag \\
&=&\frac{g^{2}}{2\pi }\int\limits_{-\pi }^{\pi }dk[\mathcal{P}\frac{1}{p-(
\frac{\Omega }{2}-\omega _{k})}\mp i\pi \delta (p-\frac{\Omega }{2}+\omega
_{k})]  \notag \\
&=&\frac{g^{2}}{2\pi }[\int\limits_{-\pi }^{\pi }dk\mathcal{P}\frac{1}{p-(
\frac{\Omega }{2}-\omega _{k})}  \notag \\
&&\mp i2\pi \int\limits_{\omega _{0}-2\zeta }^{\omega _{0}+2\zeta
}\left\vert \frac{dk}{d\omega _{k}}\right\vert d\omega _{k}\delta (p-\frac{
\Omega }{2}+\omega _{k})]  \notag \\
&=&\frac{g^{2}}{2\pi }[\int\limits_{-\pi }^{\pi }dk\mathcal{P}\frac{1}{p-(
\frac{\Omega }{2}-\omega _{k})}  \notag \\
&&\mp \int\limits_{\omega _{0}-2\zeta }^{\omega _{0}+2\zeta }\frac{i2\pi }{
2\zeta \left\vert \sin k\right\vert }d\omega _{k}\delta (p-\frac{\Omega }{2}
+\omega _{k})]  \notag \\
&=&\frac{g^{2}}{2\pi }[\int\limits_{-\pi }^{\pi }dk\mathcal{P}\frac{1}{p-(
\frac{\Omega }{2}-\omega _{k})}  \notag \\
&&\mp \int\limits_{\omega _{0}-2\zeta }^{\omega _{0}+2\zeta }\frac{i2\pi
\delta (p-\frac{\Omega }{2}+\omega _{k})}{\sqrt{(2\zeta )^{2}-(\omega
_{0}-\omega _{k})^{2}}}d\omega _{k}]  \notag \\
&=&\frac{g^{2}}{2\pi }[\int\limits_{-\pi }^{\pi }dk\mathcal{P}\frac{1}{p-(
\frac{\Omega }{2}-\omega _{k})}\mp \frac{i2\pi }{\sqrt{(2\zeta )^{2}-(\omega
_{0}+p-\frac{\Omega }{2})^{2}}}]\text{,}  \notag \\
&&  \label{DENOM}
\end{eqnarray}
where the principal value function
\begin{equation}
\mathcal{P}\frac{1}{p-(\frac{\Omega }{2}-\omega _{k})}=\left\{
\begin{array}{c}
0 \\
\frac{1}{p-(\frac{\Omega }{2}-\omega _{k})}%
\end{array}
\right.
\begin{array}{c}
\text{if }p-(\frac{\Omega }{2}-\omega _{k})=0 \\
\text{if }p-(\frac{\Omega }{2}-\omega _{k})\neq 0%
\end{array}
\text{,}
\end{equation}
and the additional factor $2$ in the fourth line is due to the same
contribution from $\pm k$.

And the first term on the right hand side of Eq.~(\ref{DENOM})
\begin{eqnarray}
\int\limits_{-\pi }^{\pi }dk\mathcal{P}\frac{1}{p-(\frac{\Omega }{2}-\omega
_{k})} &=&\int\limits_{-\pi }^{\pi }dk\mathcal{P}\frac{1}{p-\frac{\Omega }{2}
+\omega _{0}-2\zeta \cos k}  \notag \\
&=&\oint\limits_{\left\vert z\right\vert =1}\frac{dz}{iz}\mathcal{P}\frac{1}{
(p-\frac{\Omega }{2}+\omega _{0})-\zeta (z+\frac{1}{z})}  \notag \\
&=&\oint\limits_{\left\vert z\right\vert =1}\frac{dz}{-i\zeta }\mathcal{P}
\frac{1}{z^{2}-\frac{p-\frac{\Omega }{2}+\omega _{0}}{\zeta }z+1}  \notag \\
&=&\frac{2\pi i}{-i\zeta }\sum \mathrm{res[}\frac{1}{z^{2}-\frac{p-\frac{
\Omega }{2}+\omega _{0}}{\zeta }z+1}]\text{,}  \notag \\
&&
\end{eqnarray}
where the summation is over the residue within the loop enclosed by $%
\left\vert z\right\vert =1$. Apparently, there are two solutions to the
equation
\begin{equation}
z^{2}-\frac{p-\frac{\Omega }{2}+\omega _{0}}{\zeta }z+1=0\text{,}
\end{equation}
i.e.,
\begin{equation}
z_{\pm }=\frac{M_{1}\pm i\sqrt{4-M_{1}^{2}}}{2}
\end{equation}
with
\begin{equation}
M_{1}=\frac{p-\frac{\Omega }{2}+\omega _{0}}{\zeta }\text{.}
\end{equation}
For the branch cut
\begin{equation}
p\in \left[\Omega/2-\omega _{0}-2\zeta ,\Omega/2-\omega _{0}+2\zeta\right]%
\text{,}
\end{equation}
we have
\begin{eqnarray}
M_{1} &\in &[-2,2]\text{,} \\
\left\vert z_{\pm }\right\vert &=&1\text{.}
\end{eqnarray}
Furthermore, due to the principal value function $\mathcal{P}$, the above
two singular points are removed from the integral path. As a result,
\begin{equation}
\int\limits_{-\pi }^{\pi }dk\mathcal{P}\frac{1}{p-(\frac{\Omega }{2}-\omega
_{k})}=0\text{.}
\end{equation}
Then, by substituting the above equation into Eq.~(\ref{DENOM}), we have
\begin{equation}
\sum\limits_{k}\frac{g_{k}^{2}}{p-(\frac{\Omega }{2}-\omega _{k})\pm i0^{+}}
=\mp \frac{ig^{2}}{\sqrt{(2\zeta )^{2}-(\omega _{0}+p-\frac{\Omega }{2})^{2}}
}\text{.}
\end{equation}
Therefore, the contribution from the branch cut can be further simplified as
\begin{equation}
-\int\limits_{l_{1}}-\int\limits_{l_{2}}=\int\limits_{-2\zeta }^{2\zeta
}C(x)e^{i(\frac{\Omega }{2}-\omega _{0}+x)t}dx
\end{equation}
with
\begin{equation}
C(x)=\frac{1}{\pi }\frac{g^{2}\sqrt{4\zeta ^{2}-x^{2}}}{(4\zeta
^{2}-x^{2})(\Omega -\omega _{0}+x)^{2}+g^{4}}\text{.}  \label{Cx}
\end{equation}

In conclusion, the final solution is written as
\begin{eqnarray}
\alpha (t) &=&\sum_{j}\mathrm{res}[\alpha
(p_{j})e^{p_{j}t}]-\int\limits_{l_{1}}-\int\limits_{l_{2}}  \notag \\
&=&A_{1}e^{p_{1}t}+A_{2}e^{p_{2}t}+\int\limits_{-2\zeta }^{2\zeta }C(x)e^{i(%
\frac{\Omega }{2}-\omega _{0}+x)t}dx\text{,}  \notag \\
&&
\end{eqnarray}
\newline
where the coefficients $A_{1}$, $A_{2}$, and $C(x)$ are real as given in
Eqs.~(\ref{A1}),~(\ref{A2}), and~(\ref{Cx}) respectively. Here, two pure
image numbers $p_{j}$are the solutions to
\begin{equation}
\left(ip-\frac{\Omega }{2}\right)^{2}\left[\left(ip+\frac{\Omega }{2}-\omega
_{0}\right)^{2}-4\zeta ^{2}\right]+g^{4}=0
\end{equation}
with
\begin{equation}
ip_{1}+\frac{\Omega }{2}-\omega _{0}>2\zeta,
\end{equation}
and
\begin{equation}
ip_{2}+\frac{\Omega }{2}-\omega _{0}<-2\zeta.
\end{equation}

The decay rate without measurement is defined as
\begin{equation}
R(t)=-\frac{d\left\vert \alpha (t)\right\vert ^{2}}{dt}/\left\vert \alpha
(t)\right\vert ^{2}\text{,}
\end{equation}
where the survival probability for the initial state is
\begin{equation}
\left\vert \alpha (t)\right\vert ^{2}=\left\vert
A_{1}e^{p_{1}t}+A_{2}e^{p_{2}t}+\int\limits_{-2\zeta }^{2\zeta }C(x)e^{i(%
\frac{\Omega }{2}-\omega _{0}+x)t}dx\right\vert ^{2}
\end{equation}
and its rate of change is
\begin{eqnarray}
\frac{d\left\vert \alpha (t)\right\vert ^{2}}{dt} &=&2\Re (I_{1}\times
I_{2}^{\ast })  \notag \\
&&-2A_{1}A_{2}(ip_{1}-ip_{2})\sin (ip_{1}-ip_{2})t  \notag \\
&&+2\Re \lbrack (A_{1}p_{1}e^{p_{1}t}+A_{2}p_{2}e^{p_{2}t})I_{1}^{\ast }]
\notag \\
&&+2\Re \lbrack (A_{1}e^{p_{1}t}+A_{2}e^{p_{2}t})I_{2}^{\ast }],
\end{eqnarray}
where the integrals are defined as
\begin{eqnarray}
I_{1} &=&\int\limits_{-2\zeta }^{2\zeta }C(x)e^{i(\frac{\Omega }{2}-\omega
_{0}+x)t}dx, \\
I_{2} &=&\int\limits_{-2\zeta }^{2\zeta }C(x)e^{i(\frac{\Omega }{2}-\omega
_{0}+x)t}i(\frac{\Omega }{2}-\omega _{0}+x)dx,
\end{eqnarray}
and $\Re (x)$ is the real part of $x$.

\end{document}